\documentclass{iaus}
\usepackage{graphicx} 

\title[Deriving AGN properties from radio CP and LP] 
{Deriving AGN properties from circular and linear radio polarimetry}

\author[E. Cenacchi, A. Kraus \& K.-H. Mack]   
{Elena Cenacchi$^1$%
  \thanks{E.C. is a member of the International Max-Planck Research School for Radio and Infrared Astronomy},
 A. Kraus$^1$ \break \and K.-H. Mack$^2$}

\affiliation{$^1$Max-Planck-Institut f\"ur Radioastronomie, Auf dem H\"ugel 69, D-53121 Bonn, Germany \break email: cenacchi@mpifr.de, akraus@mpifr.de\\[\affilskip]
$^2$Istituto di Radioastronomia, INAF, Via P. Gobetti 101, I-40129 Bologna, Italy \break email: mack@ira.inaf.it}

\pubyear{2009}
\volume{259}  
\pagerange{...}
\date{??}
\jname{Proceedings Title IAU Symposium}
\editors{K. G. Strassmeier, A. G. Kosovichev \& J. Beckmann, eds.}

\begin{document}
\maketitle

\begin{abstract}
We report multi-frequency circular polarization measurements for the radio source 0056-00 taken at the Effelsberg 100-m radiotelescope. The data reduction is based on a new calibration procedure that allows the contemporary measurement of the four Stokes parameters with single-dish radiotelescopes.
\end{abstract}
\vspace{-0.2cm}
\section{Introduction}   
Multi-frequency full Stokes polarimetry is a powerful tool to study the radiation emission and transfer processes in the observed sources and 
determine the dominant mechanism for circular polarization (CP) production. CP, linear polarization (LP) and spectral information can be used to constrain the low energy end of the relativistic particle distribution (\cite{Beckert}), derive magnetic field strength and geometry (\cite{Gabuzda}) and make assumptions about the composition of the relativistic plasma within jets (\cite{Wardle}). 

We have developed a new procedure to calibrate full-Stokes polarimetric data obtained with the Effelsberg 100-m radiotelescope (\cite{Cenacchi}) and have applied it at 2.7, 5, 8.5 and 10-GHz.

\begin{figure}[!t]
\label{p:0056-00}
\includegraphics[width=320 pt]{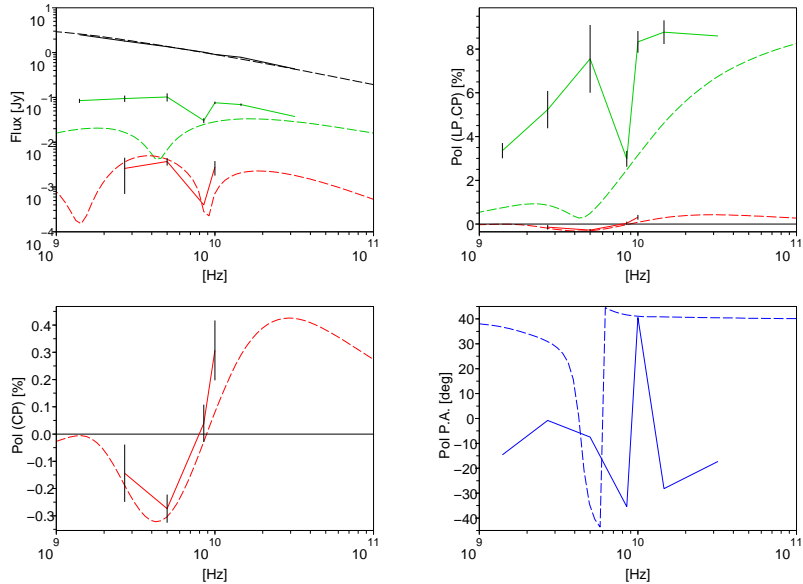}
\caption{0056-00, each plot contains a comparison of the observed radio continuum spectra (continuum line) and the simulated one (dashed line). Upper left, top to bottom: flux density I (the predicted values are almost coincident with the observed one), LP, CP (absolute value). Upper right, top to bottom: LP, CP. Lower left: CP. Lower right: PA.}
\end{figure}
\vspace{-0.2cm}
\section{Results}
Here we present, as an example, the results of our observations on 0056-00. This source, point-like to the Effelsberg beams, has been observed cross-scans, resulting in the simultaneous measurement of all four Stokes parameters. The data were reduced using suited calibrators to correct the 4$\times$4 M\"uller matrix.
0056-00 exhibits a change in sign of CP between 5 and 8.5 GHz, consistent with the minimum in LP and with the change in polarization angle. We have compared our results with those obtained with a model developed by Beckert \& Falcke (2002) that provides the radiative transfer coefficients for polarized synchrotron radiation applied to the standard model for relativistic radio jets. The model assumes an extended unpolarized synchrotron source, which dominates the flux density below nearly 5 GHz which is modeled with energy equipartition between B-field and particles, $B$=4 mG, $n_e=10^{-3}$ cm$^{-3}$, and  a typical power-law for electrons above $\gamma$=100, with $p$=2.45 (power-law index). The size of the emitting region is $L=8 \cdot 10^{21}$ cm. The polarized emission would be produced by a compact jet component of $L=1.5\cdot 10^{19}$ cm, $B$=90 mG, $n_e$=0.5 cm$^{-3}$ with a well-ordered magnetic field (a tightly wound spiral) seen at an angle of 85$^\circ$. This component becomes self-absorbed below 6 GHz and the emission is relativistically boosted with $\Gamma$=6. This combination reproduces the observed level of circular polarization, the sign flip at nearly 8 GHz and the observed flux density (Fig. 1). With respect to the observed behaviour, the minimum in LP is at too low a frequency. Also the overall level of the simulated LP is too low, which indicates that there might be an additional component (even more compact) that dominates LP and produces the turn in polarization angle at higher frequencies. The observed levels of CP are: $(-0.14\pm 0.11\%)$ at 2.8 GHz, $(-0.28\pm 0.05\%)$ at 5 GHz, $(0.04\pm0.07\%)$ at 8.5 GHz, $(0.31\pm 0.11\%)$ at 10.GHz.\\

To summarize, the comparison between the observed polarimetric parameters and the modeled simulation shows that the values observed in 0056-00 are consistent with those predicted by the model, that is currently under refinement. Further full Stokes measurements are planned at 14.5 and 32 GHz.
\vspace{-0.2cm}
\acknowledgements 
This research was supported by the EU Framework 6 Marie Curie Early Stage Training programme under contract number MEST-CT-2005-1966 ``ESTRELA''.


\vspace{-0.2cm}	
\begin{thebibliography}{}
\bibitem [Beckert 2003]{Beckert}
Beckert, T. 2003, Ap\&SS, 288, 123
\bibitem [Beckert \& Falcke 2002] {BeckertFalcke} 
Beckert, T. \& Falcke, H. 2002, A\&A, 388, 1106
\bibitem [Cenacchi et al. 2008] {Cenacchi}
Cenacchi, E., Kraus, A., Orfei, A. \& Mack K.-H. 2008, A\&A (subm.).
\bibitem [Gabuzda 2006]{Gabuzda}
Gabuzda, D. 2006, in Proceedings of the 8th European VLBI Network Symposium, Marecki, A. et al. (ids.)
\bibitem [Wardle et al. 1998]{Wardle}
Wardle, J. F. C., Homan, D. C., Ojha, R., \& Roberts, D. H. 1998, Nature, 395, 457
\bibitem [Wardle \& Homan 2003]{WardleHoman2003}
Wardle, J. F. C., \& Homan, D. C. 2003, Ap\&SS, 288, 143
\end{thebibliography}
\end{document}